\documentclass[pdflatex,sn-mathphys-num]{sn-jnl}

\usepackage{graphicx}%
\usepackage{multirow}%
\usepackage{amsmath,amssymb,amsfonts}%
\usepackage{amsthm}%
\usepackage{mathrsfs}%
\usepackage[title]{appendix}%
\usepackage{xcolor}%
\usepackage{textcomp}%
\usepackage{manyfoot}%
\usepackage{booktabs}%
\usepackage{algorithm}%
\usepackage{algorithmicx}%
\usepackage{algpseudocode}%
\usepackage{listings}%
\usepackage{anyfontsize}

\begin{document}

\title[Article Title]{Tunable microwave frequency synthesis with optically-derived spectral purity}
\author*[1]{\fnm{James} \sur{Greenberg}}\email{jgreenbe@imra.com}
\equalcont{These authors contributed equally to this work.}

\author[1]{\fnm{Scott C.} \sur{Egbert}}\email{segbert@imra.com}
\equalcont{These authors contributed equally to this work.}

\author[1]{\fnm{William F.} \sur{McGrew}}\email{wmcgrew@imra.com}

\author[1]{\fnm{Brendan M. } \sur{Heffernan}}\email{bheffern@imra.com}

\author[1]{\fnm{Antoine} \sur{Rolland}}\email{arolland@imra.com}

\affil*[1]{\orgdiv{Boulder Research Labs}, \orgname{IMRA America, Inc.}, \orgaddress{\street{1551 S. Sunset St. Suite C}, \city{Longmont}, \postcode{80501}, \state{Colorado}, \country{USA}}}

\abstract{Microwave synthesizers are central to test and measurement systems across applications including wireless communications, radar, spectroscopy, and time and frequency metrology. State-of-the-art microwave sources, however, are fundamentally constrained by trade-offs between frequency tunability and spectral purity. Electro-optic frequency division (eOFD) is an emerging technique for dividing down the purity of optical sources to the microwave domain. Previously reported eOFD-based synthesizers generally have limited tunability due to feedback stabilization requirements. Here we demonstrate a feed-forward eOFD architecture in which the frequency tunability of a microwave source is preserved while optical spectral purity is divided through feed-forward cancellation, without any downstream electronic frequency synthesis. By canceling the phase noise of the  microwave source without feedback, this eOFD approach removes loop bandwidth and source noise constraints observed in prior eOFD architectures. We achieve octave-spanning tunability, including the entire X-band, with phase noise below -140\,dBc/Hz at kilohertz offsets and a high-frequency noise floor between -155\,dBc/Hz and -145\,dBc/Hz for carrier frequencies from 8--16\,GHz. This performance corresponds to single-femtosecond integrated timing jitter, enabling, to our knowledge, the first demonstration of coherent, optically referenced microwave synthesis under wide tuning with this level of spectral purity.
}


\maketitle

Microwaves are ubiquitous among modern society, permeating many realms of technology including cellular wireless~\cite{robertson2016}, radar~\cite{skolnik2008}, atomic clocks~\cite{townes1951}, and global positioning systems~\cite{kaplan2017}. The core of such systems is a microwave synthesizer providing the steady rhythm to which everything is synchronized. In practice, microwave frequency synthesis is characterized by a trade-off between tunability and spectral purity: architectures that support wide frequency tuning rely on active electronic control and modulation stages that introduce phase noise~\cite{rohde2021}, whereas architectures that achieve exceptional spectral purity do so by strongly constraining the oscillator frequency through passive resonators with high quality factor, such as sapphire microwave cavities~\cite{ivanov2009}. This trade-off is evident even in the highest-performance voltage-controlled oscillators, such as yttrium iron garnet (YIG) oscillators. These offer exceptionally wide tuning ranges and low phase noise among tunable microwave sources~\cite{stein2017,vandelden2019}, but still fall short of the spectral purity achievable with fixed-frequency oscillators.

The most spectrally pure microwaves yet demonstrated are derived from optical frequency division (OFD), where oscillators disciplined by the extremely high quality factors available in the optical domain are divided down to the microwave regime~\cite{ye2000,bartels2005,zhang2010}. There are three distinct architectures for OFD that we will highlight. First, is full-OFD, in which a self-referenced optical frequency comb divides down the phase stability of a single optical carrier (e.g. stabilized laser) to the microwave regime ~\cite{fortier2011}. While capable of zeptosecond timing noise~\cite{xie2017}, full-OFD produces a fixed microwave frequency and requires lab-grade metrology equipment, such as an ultra-stable optical reference cavity and a self-referenced optical frequency comb, limiting applications beyond precision metrology. 

Second, is two-point OFD, in which an optical frequency comb is locked simultaneously to two optical carriers, dividing the differential phase stability down to the microwave domain~\cite{kwon2022,kudelin2024a}. While an optical frequency comb is still required, it no longer needs to be self-referenced, reducing system complexity. Furthermore, common mode noise rejection of the dual optical carriers removes the need for an ultra-stable optical reference cavity. This architecture opens the door for compact, even integrated-photonic-based demonstrations~\cite{sun2024,sun2025a,zhao2024}. However, the tunability is tightly constrained by the comb repetition rate (typically less than 1\%). Recent work with direct digital synthesis and single-sideband mixing has partially relaxed this constraint by adopting downstream electronic synthesis techniques~\cite{nand2011,fortier2016,kudelin2024}. Because the frequency agility is introduced after optical-to-microwave conversion, phase-noise performance away from the carrier is ultimately limited by the electronic synthesis stage, removing some of the spectral purity benefits of OFD.

The third architecture is electro-optic frequency-division (eOFD). Like two-point OFD, a pair of optical carriers is again used as the reference. Instead of comparing the two references through an external optical frequency comb, eOFD uses an electro-optic modulator to generate a comb from a microwave source that spans the optical frequency difference, carying the phase information of both electrical and optical oscillators. One of the principal advantages of this technique is arbitrary choice of microwave frequency used for the optical modulation. Dividing the optical difference stability is then typically accomplished via feedback on the microwave source, requiring intrinsically low-noise oscillators that support wideband phase actuation\cite{li2014,he2024,egbert2025a}. In practice, this generally restricts operation to dielectric resonator oscillators (DROs) or sapphire-resonator stabilized sources. While these oscillators offer favorable noise properties, they also provide limited tuning ranges, typically below 5\% of the carrier frequency, and necessitate control loops with bandwidths approaching or exceeding 1 MHz. Fully agile microwave synthesizers cannot be used in feedback-based eOFD, as their intrinsic phase noise and limited actuation bandwidth prevent the feedback loop from reaching the noise floor set by the optical reference. As a result, spectral purity and wideband tunability remain incompatible in feedback-stabilized eOFD systems.

Feed-forward noise cancellation provides an alternative to feedback-based stabilization. Nakamura et al. demonstrated electrical feed-forward correction to suppress repetition-rate noise in free-running optical frequency combs, removing loop dynamics and avoiding servo-induced artifacts~\cite{nakamura2025}. However, this approach relied on fixed repetition-rate combs (two-point OFD) and produced fixed microwave outputs, leaving the constraint on frequency tunability unaddressed.

Here we introduce a feed-forward electro-optic frequency-division (eOFD) synthesizer that divides the spectral purity of the difference of two highly correlated optical tones to a free-running, tunable microwave source without feedback stabilization. An optical reference spacing is compared to an electro-optically multiplied microwave signal to generate a beatnote encoding the instantaneous oscillator phase error, which is processed and recombined to cancel phase noise algebraically at the output. Because no feedback is applied, the approach imposes no loop bandwidth or actuation constraints and does not require a low-noise, controllable oscillator. Using a dielectric resonator oscillator, we show that feed-forward eOFD surpasses feedback-based architectures at Fourier frequencies from 100~kHz to 10~MHz by eliminating servo-induced noise peaking, achieving phase noise of $-150~\mathrm{dBc/Hz}$ at 10~kHz offset and $-162~\mathrm{dBc/Hz}$ at 100~kHz. When applied to a fully tunable microwave synthesizer, the output can be changed to frequencies from 8 to 16~GHz while maintaining optical coherence, with phase noise below $-140~\mathrm{dBc/Hz}$ at kilohertz offsets and a high-frequency noise floor between $-155~\mathrm{dBc/Hz}$ and $-145~\mathrm{dBc/Hz}$, depending on the carrier frequency. This corresponds to single femtosecond integrated timing jitter, which is a factor of 10 lower than state-of-the-art agile microwave sources. Feed-forward eOFD decouples spectral purity from oscillator noise and control bandwidth, establishing a synthesis architecture in which optical references define spectral purity and electronic oscillators define tunability.

\section*{Feed-forward electro-optic frequency-division architecture}

\begin{figure}[htp]
\centering
\includegraphics[width=0.95\textwidth]{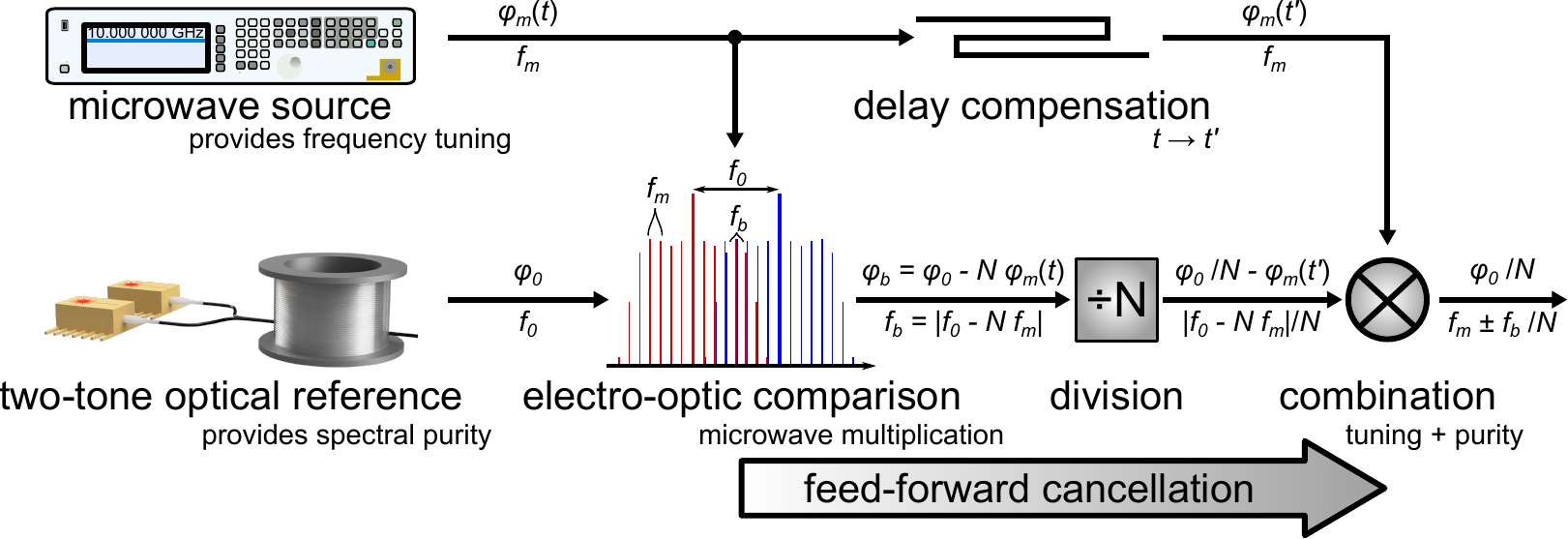}
\caption{\textbf{eOFD with feed-forward concept.} The phase and frequency of a microwave oscillator denoted by $\varphi_m$ and $f_m$, respectively. They are tracked above and below the arrows through the conceptual diagram. Meanwhile, the phase and frequency difference of a two-tone, optical reference are denoted by $\varphi_0$ and $f_0$, respectively. Electro-optical modulation allows the microwave signal to be multiplied by an integer $N$, up to the optical frequency difference and compared. The phase difference is carried by a beat-note $f_b$. Subsequent division of the beat-note and combination with a delayed copy of the original microwave source produces the output of the oscillator. The output phase is determined by the divided optical frequency reference phase, while the output frequency is mostly determined by the input microwave source frequency.}\label{fig:concept}
\end{figure}

Figure~\ref{fig:concept} illustrates the operating principle of feed-forward eOFD, highlighting how the architecture explicitly separates the roles of the microwave and optical domains. A free-running microwave source provides frequency tunability, while an optical frequency difference serves as a spectrally pure phase reference.

The microwave source, with instantaneous phase $\varphi_m(t)$ at carrier frequency $f_m$, is split into two paths. In the lower path, the microwave signal is applied to an electro-optic modulator, generating a comb of sidebands around each optical tone, spaced by $f_m$. This electro-optically generated comb enables a direct comparison between the multiplied microwave frequency $Nf_m$ and the optical reference spacing
$f_0$. Photodetection produces an intermediate-frequency beat note
\begin{equation}
f_b = |f_0 - N f_m|,
\end{equation}
with phase encoding the instantaneous phase difference
\begin{equation}
\varphi_b(t) = \varphi_0 - N \varphi_m(t),
\end{equation}
where $\varphi_0$ is the phase of the optical reference.

The beat signal is subsequently divided by $N$, yielding a correction signal with phase $\varphi_0/N - \varphi_m(t')$, where $t'-t$ is the time delay between splitting the paths and recombining at the mixer. In the upper path, the original microwave signal propagates through a delay-compensation stage, introducing a relative time shift $t \rightarrow t''$. The divided correction signal is then combined with this delayed microwave signal. In the ideal case of perfect delay matching ($t' = t''$), the microwave phase $\varphi_m$ cancels algebraically, and the output phase is set by the divided optical reference,
\begin{equation}
\varphi_{\mathrm{out}} = \frac{\varphi_0}{N}.
\end{equation}

Because the phase correction is applied in a feed-forward manner, no feedback is imposed on the microwave source. As a result, the source remains free running and tunable, while the output inherits the spectral purity of the optical reference. Thus, tunability and purity are combined without constraints from servo bandwidth or actuation range.

Highlighting the simplicity of the architecture, Eq.~\eqref{eq:pnmodel} contains a single-sideband phase-noise power spectral density model capturing all residual noise contributions, as derived in the Methods section. 
\begin{equation}
S^{\mathrm{out}}_{\phi}(f) = \frac{1}{N^2}S^{0}_{\phi}(f)
+ \frac{1}{N^2}S^{b}_{\phi}(f)
+ S^{\mathrm{div}}_{\phi}(f)
+ 4\sin^2(\pi f\Delta\tau)~S^{m}_{\phi}(f),
\label{eq:pnmodel}
\end{equation}
where $S^{0}_{\phi}$ is the optical reference phase-noise PSD, $S^{b}_{\phi}$ accounts for additive noise in the optical beat detection, $S^{\mathrm{div}}_{\phi}$ represents the residual noise of the frequency-division electronics, $S^{m}_{\phi}$ is the phase noise of the microwave source, and $\Delta\tau = t - t'$ is the delay mismatch between the two microwave paths.

Equation~\eqref{eq:pnmodel} highlights that, in the absence of delay mismatch, the microwave source phase noise is fully suppressed at the output. A finite delay mismatch introduces a frequency-dependent residual contribution proportional to the source phase noise, which increases with Fourier frequency. This behavior is intrinsic to feed-forward cancellation and establishes a clear engineering requirement on path-length matching, with stringency determined by the phase-noise level of the microwave source. Experimental validation of this
dependence for multiple oscillators is provided in Extended Data Fig. \ref{fig:sources}.

\section*{Results}\label{sec:results}

\begin{figure}[htp]
\centering
\includegraphics[width=0.95\textwidth]{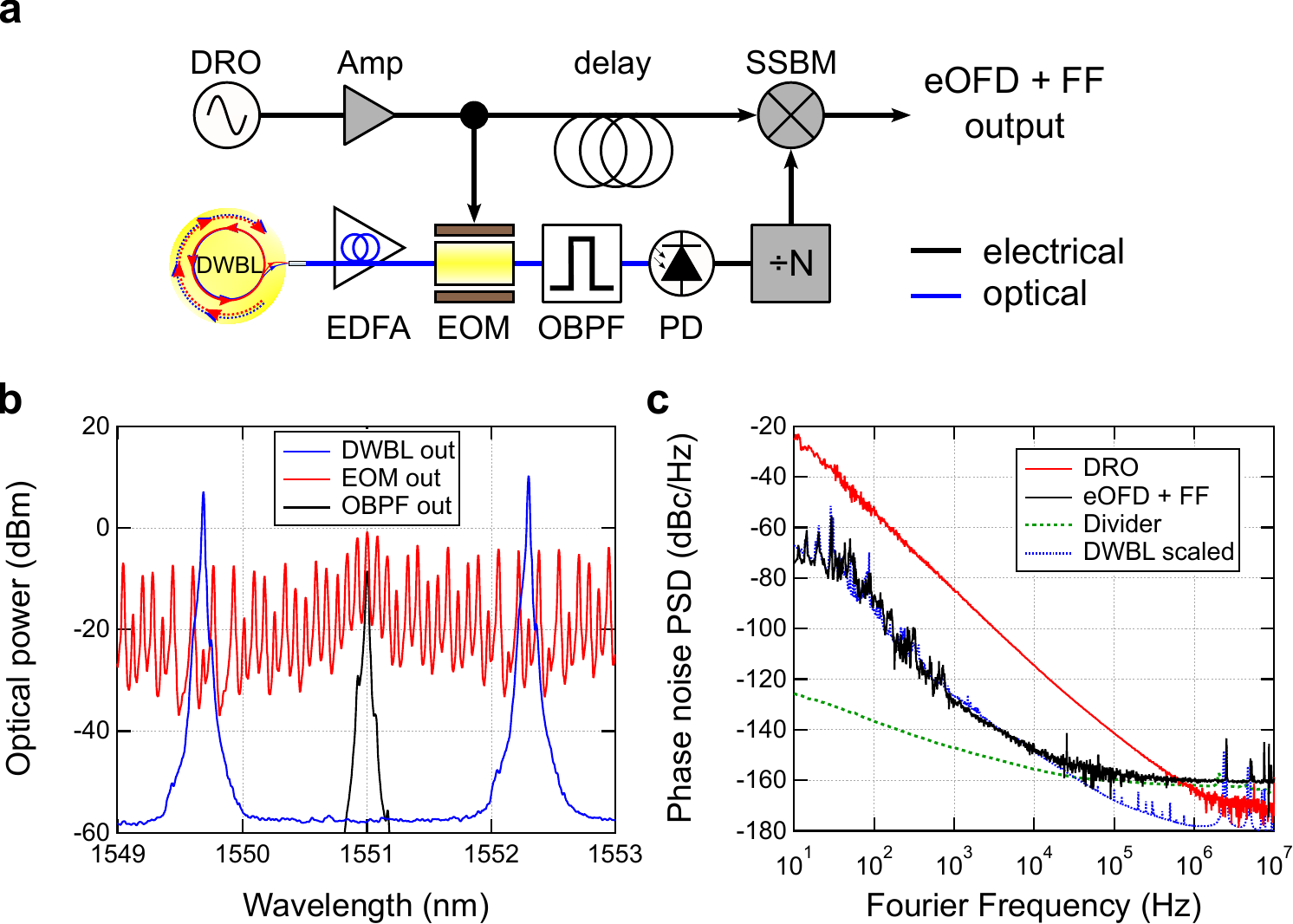}
\caption{\textbf{eOFD with feed-forward schematic and output phase noise.} \textbf{a} Schematic used for experimental demonstration of eOFD with feed-forward. DWBL: dual-wavelength Brillouin laser, EDFA: erbium-doped fiber amplifier, EOM: electro-optic modulator, OBPF: optical band-pass filter, PD: photodiode, DRO: dielectric resonator oscillator, SSBM: single-sideband mixer. \textbf{b} Optical power spectra along the lower arm of the eOFD path. \textbf{c} Phase noise power spectral density (PSD) of the 10~GHz input and output after feed-forward. The scaled optical reference (DWBL) and divider residual floor are also shown, which provide the limits for the output phase noise.}\label{fig:eOFDFF}
\end{figure}

To experimentally validate the feed-forward electro-optic frequency-division architecture and phase-noise model introduced in the previous section, we first implemented the system using a low-noise dielectric resonator oscillator (DRO). This configuration represents a best-case scenario in which the microwave source exhibits low phase noise at high Fourier frequencies, allowing the feed-forward cancellation mechanism to be evaluated independently of source limitations from delay mismatch.

Implementing the phase noise cancellation of eOFD with feed-forward under these conditions, we utilized the schematic in Fig. \ref{fig:eOFDFF}\textbf{a}. A dual wavelength Brillouin laser (DWBL) with an optical difference of $f_0=320$~GHz was used as the optical reference. The DWBL has been described in detail previously \cite{heffernan2024,egbert2025a}. Briefly, two diode lasers non-resonantly pumped a common, non-reciprocal fiber cavity. The resulting stimulated Brillouin scattering (Stokes wave) was used both as the laser output and to injection lock each diode laser to the fiber cavity. The optical frequency difference between the Stokes waves was discretely tunable with megahertz-scale resolution over the gigahertz to terahertz range and exhibited strongly correlated phase noise that was common-mode rejected upon photodetection.

A DRO oscillating at 10~GHz (Synergy, DRO100) was used as the microwave oscillator, along with a low-residual-noise power amplifier. A coupler was placed at the output of the amplifier with the majority of the microwave power being used to modulate the DWBL light via an electro-optic phase modulator (Hyperlight, TFLN-G2-PM-1550-ULV). The modulated light was optically band-pass filtered with a narrow $<10$~GHz bandwidth. The filtered light was detected by a photodiode producing $f_b\approx1$~GHz (arbitrary), which carried the phase information of the DWBL and the DRO multiplied by $N=32$. Fig. \ref{fig:eOFDFF}\textbf{b} shows these optical signals at the output of the DWBL, after the EOM, and immediately before photodetection.

After photodetection, $f_b$ was divided by $N=32$ (Valon, 3010a) and sent to the IF port of a single-sideband RF mixer (see Methods). The rest of the amplified DRO output was sent through a delay line followed by the LO port of the single-sideband mixer. Delay  optimization is described in the methods section. The compensated output was then analyzed by a phase noise analyzer equipped with dual internal references and cross-correlation (Keysight SSA-X). 

Fig. \ref{fig:eOFDFF}\textbf{c} shows the single-sided phase noise PSD of the output 10~GHz signal along with the phase noise of the optical reference scaled by the division ratio of~32, the DRO, and the frequency divider noise floor. The observed suppression of the DRO phase noise and the absence of servo-induced features are consistent with the feed-forward phase-noise model described in Eq.~\eqref{eq:pnmodel}. The DRO phase noise PSD was improved to the level of the scaled DWBL at Fourier frequencies below 50~kHz. Above 50~kHz, the output is limited by the residual phase noise floor of the frequency divider at -162~dBc/Hz. No servo bump is present in the output phase noise as no electronic feed-back or servos were necessary. The resulting integrated timing jitter (1\,khz to 10\,MHz) was below one femtosecond (See Extended Data Fig. \ref{fig:inttj}).

While the DRO-based implementation establishes the fundamental operation and ultimate phase-noise limits of feed-forward eOFD, it does not address the central constraint of conventional microwave synthesis, namely the tradeoff between spectral purity and frequency tunability. Dielectric resonator oscillators are inherently narrowband and do not support the wide tuning required in many electronic systems. The key promise of the feed-forward architecture is that phase-noise suppression does not rely on feedback or on the intrinsic noise properties of the microwave source. 

\subsection*{Tunable frequency synthesis}\label{subsec:tunable freq}

\begin{figure}[htp]
\centering
\includegraphics[width=0.80\textwidth]{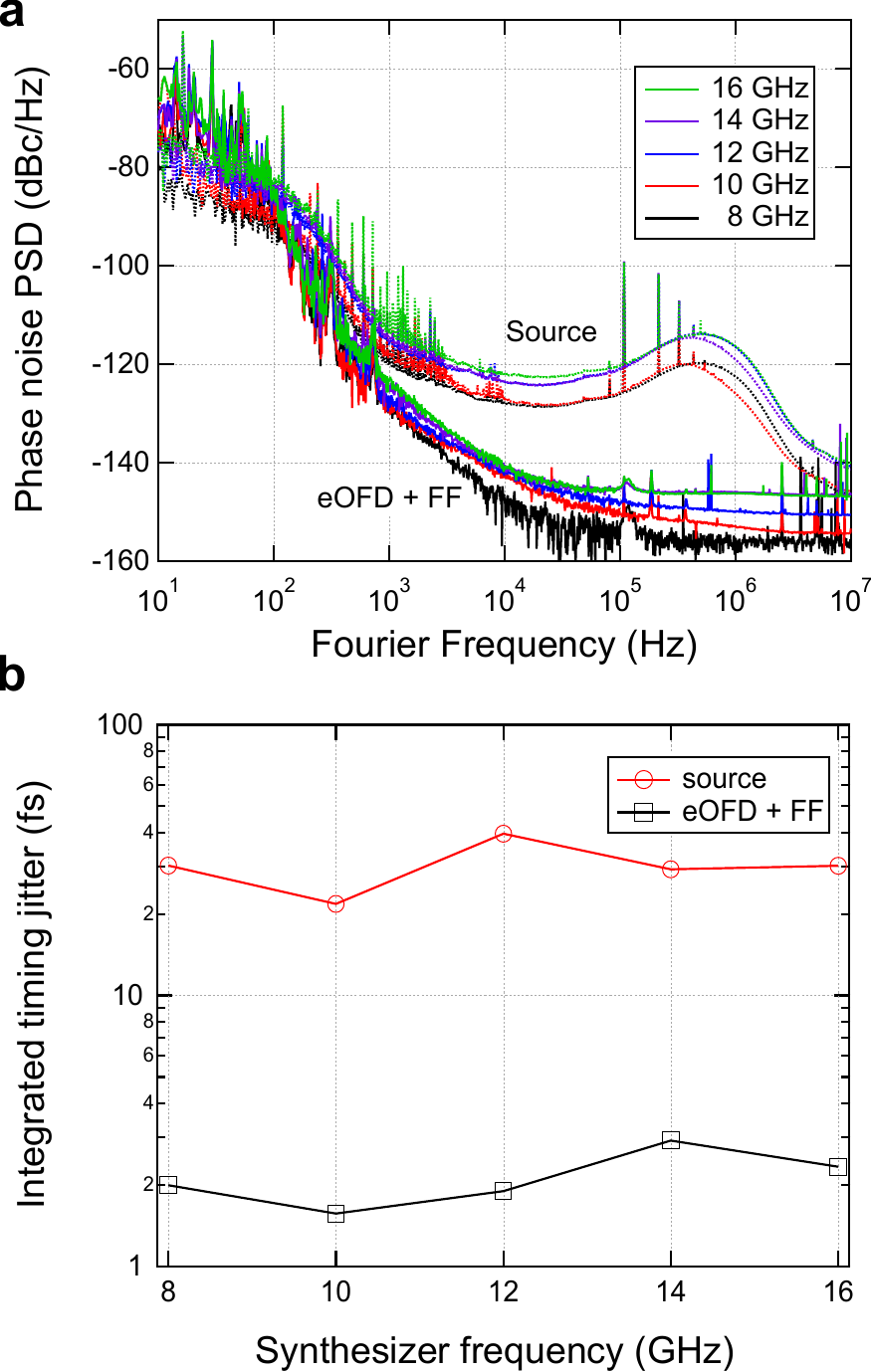}
\caption{\textbf{Synthesizer phase noise at different output frequencies and corresponding integrated timing jitter. a} Phase noise PSD before and after eOFD with feed-forward for several output frequencies. The dotted traces are the microwave source phase noise while the solid traces are the phase corrected outputs. Above 1~kHz Fourier frequency, the synthesizer output has a significantly lower phase noise than its corresponding microwave source input. \textbf{b} Integrated timing jitter from 1~kHz to 10~MHz as a function of output frequency. The eOFD and feed-forward technique improves the integrated timing noise by a full factor of 10 over an octave spanning output. The result is a synthesizer with single-digit femtosecond timing jitter.}\label{fig:synthsweep}
\end{figure}

We next demonstrate the central consequence of the feed-forward eOFD architecture: ultra-low-noise microwave synthesis under wideband frequency tuning. Using a fully tunable commercial microwave synthesizer as a source, we achieved octave-spanning output tunability while retaining optical reference-limited phase noise.

For this demonstration, we again used the schematic from Fig. \ref{fig:eOFDFF}\textbf{a} and $N=32$ for the multiplication and division. We maintained a constant $f_b$ and $N$, while $f_0$ varied for each output frequency (from 256~GHz to 512~GHz optical span). Using a tunable microwave source to vary $f_m$ (Keysight, MXG N5183B), we took measurements across an octave of frequencies from 8~GHz to 16~GHz. The output phase noise PSD for each of these frequencies are shown Fig. \ref{fig:synthsweep}\textbf{a} along with the phase noise of the tunable synth directly. A dramatic reduction in phase noise can be seen above 1~kHz Fourier frequency. The slight increase in phase noise at higher output frequencies is likely due to power balance between the two optical tones limiting SNR of $f_b$ at larger optical spans, with minor contributions from the increased phase noise of the microwave source through the last term of Eq. \ref{eq:pnmodel}. Together, these caused the phase noise floor to increase from -155~dBc/Hz to -145~dBc/Hz when tuning from 8 to 16~GHz. Some of this increase could be mitigated by reducing the optical reference spacing by using a smaller division ratio. While the $1/N^2$ reduction in phase noise would be smaller, improvements in $S^{b}_{\phi}(f)$ and likely $S^{div}_{\phi}(f)$ could lead to a net decrease in the total phase noise at the output. These highly coupled optimizations were not explored in this work as the relevant tradeoffs are application specific.

The integrated timing jitter as a function of output frequency is shown in Fig.~\ref{fig:synthsweep}\textbf{b}. It was obtained by integrating the measured phase noise from 1~kHz to 10~MHz for both the microwave synthesizer and the output after eOFD and feed-forward. The uncompensated synthesizer represents a high-performance, best-in-class commercial microwave source, making the observed jitter reduction particularly significant. Across the entire octave-spanning tuning range, the feed-forward eOFD output exhibits an order-of-magnitude reduction in integrated timing jitter relative to the free running synthesizer, with single-femtosecond performance maintained at all output frequencies.

Integration from 1\,kHz to 10\,MHz was chosen as it is the most relevant band for low-noise, tunable synthesizer applications (e.g., analog-to-digital conversion and radar). The full integrated timing jitter as a function of Fourier frequency is also provided in the Extended Data Fig. \ref{fig:inttj}. Below 1\,kHz offset, frequency drift becomes the dominant contribution to the integrated timing jitter. Mitigation of the frequency drift is discussed in the following section.

\subsection*{Frequency drift and absolute reference}\label{subsec:clocking}

\begin{figure}[tp]
\centering
\includegraphics[width=0.95\textwidth]{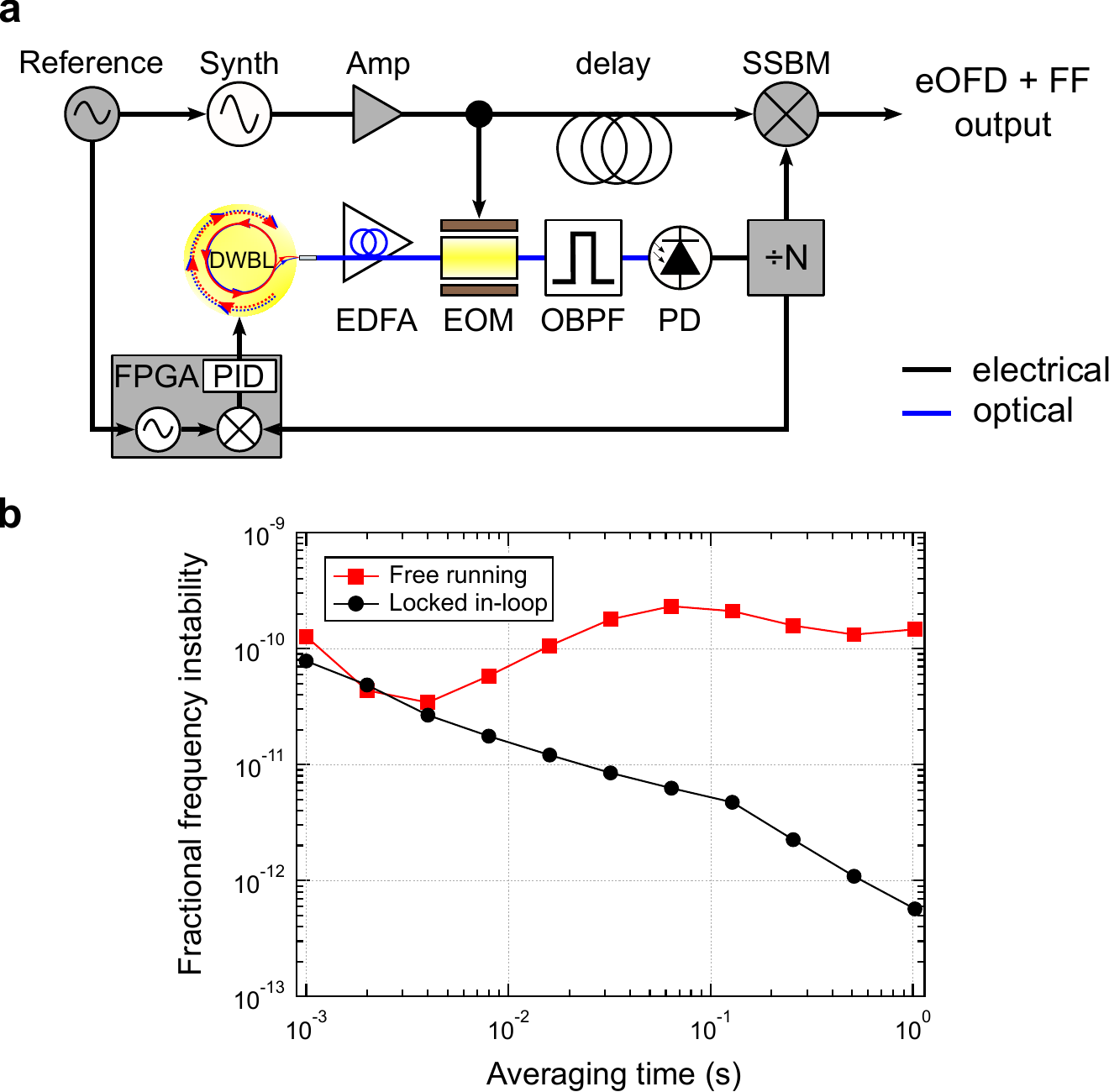}
\caption{\textbf{eOFD and feed-forward synthesizer with reference and feedback. a} Schematic of the tunable synthesizer with feedback added for referencing to an atomic frequency standard. The same beat signal used for feed-forward corrections can be used to feed back to the optical reference. FPGA: field programmable gate array, PID: proportional-integrator-derivative. \textbf{b} Residual fractional frequency instability of the synthesizer while locked to an atomic frequency reference. The synthesizer was capable of supporting a reference with $5\times10^{-13}$ fractional frequency stability at 1~s averaging time. Without referencing, the DWBL drift determines the output stability, leading to $>100$ times worse fractional frequency stability at 1~s averaging time.}\label{fig:feedback}
\end{figure}

Beyond frequency tunability and short-term spectral purity, practical microwave synthesizers must support long-term frequency stability through an external reference. In the feed-forward eOFD architecture, this functionality also addresses residual low-Fourier-frequency noise arising from drift in the spacing of the optical reference.

Previously, we have referenced DWBL oscillators to rotational transitions from a molecular frequency reference \cite{greenberg2025}. Here we show an alternative method that compares the eOFD beat signal to a standard 10~MHz atomic reference, as shown in Fig. \ref{fig:feedback}\textbf{a}, feeding back to the DWBL optical self-injection to stabilize the relative spacing of the optical lines. This method utilized the modest continuous frequency tuning range of the DWBL ($\approx 100$\,kHz) and feedback bandwidth ($\leq 1$\,kHz).

The optical beat signal $f_b$ contained the fluctuations of both the optical reference $f_0$ and the microwave reference $f_m$. After frequency division, $f_b/N$ provided an error proportional to the phase difference between $\phi_0/N$ and the microwave phase $\phi_m$. Because $\phi_m$ can be independently disciplined to the 10~MHz atomic reference, this synchronized the optical reference spacing $\phi_0$ to the same atomic clock at low Fourier frequencies. We used an FPGA-based phase-locked loop (PLL) to control the DWBL optical reference spacing. In doing so, the output fractional frequency fluctuations followed that of the atomic reference. To characterize how well the PLL can stabilize the absolute frequency of the output, we measured the output with a counter clocked by the same atomic reference. We then calculated the Allan deviation of the frequency fluctuations to report the residual instability of the fractional frequency fluctuations, as shown in Fig. \ref{fig:feedback}\textbf{b}.

The residual instability reaches the $10^{-13}$ level at a one second averaging time, sufficient for following most commercially available timing standards. The locked data is also compared to the free-running DWBL, which led to $>$100 times worse fractional frequency instability at 1~s. These data also explain why the eOFD phase noise is greater than the free-running synthesizers for low Fourier frequency (see Fig. \ref{fig:synthsweep}\textbf{a} and Extended Data Fig. \ref{fig:timedelay}). The DWBL phase noise is written to the synthesizer which can degrade the low Fourier frequency performance. These data show that less than a kilohertz bandwidth PLL using the same $f_b$ signal can remedy the drift. Together, the feed-forward architecture and low-bandwidth optical referencing enable simultaneous frequency tunability, ultralow timing jitter, and absolute long-term frequency stability within a single microwave synthesizer platform.

\section*{Conclusion}\label{sec:conclusion}

In this work, we have demonstrated a feed-forward electro-optic frequency-division architecture that enables microwave synthesis with optical-reference-limited spectral purity while preserving frequency tunability. By decoupling phase-noise suppression from microwave oscillator quality and feedback constraints, the approach achieves ultralow timing jitter using free-running and fully tunable electronic sources.

We first established the fundamental operation and noise limits of feed-forward eOFD using a low-noise but minimally tunable, dielectric resonator oscillator, validating a quantitative phase-noise model that captures all dominant residual contributions, including the role of differential path delay. Extending the architecture to a high-performance commercial microwave synthesizer, we demonstrated octave-spanning frequency tunability with an order-of-magnitude reduction in integrated timing jitter and sustained single-femtosecond performance across the tuning range. This overcomes the longstanding tradeoff in optical frequency-division techniques between unmatched spectral purity and limited tunability.

Although the present implementation provides wide tunability, it does not yet enable fully arbitrary high-speed frequency agility. Because the instantaneous output frequency is defined by the divided optical spacing $f_0/N$, dynamic modulation of the microwave output requires high-bandwidth control of the optical reference spacing. The modulation bandwidth was limited in this demonstration by the tuning dynamics of the DWBL, which was agile enough to correct its own frequency drift, but not able to provide wide-band or fast frequency sweeps relevant to some applications. Extending the architecture to support more agile operation is the subject of future investigation, requiring high-bandwidth optical spacing control while preserving feed-forward phase-noise cancellation.

Beyond frequency tunability and short-term spectral purity, we showed that the synthesizer can be referenced to an atomic frequency standard by applying a low-bandwidth feedback loop to the optical reference spacing, mitigating long-term drift and low-Fourier-frequency noise. This hybrid feed-forward and feedback architecture enables simultaneous frequency tunability, ultralow short-term timing jitter, and absolute long-term frequency stability within a single microwave synthesizer platform.

One immediate implication is in high-speed analog-to-digital conversion, where sampling performance is fundamentally limited by clock timing jitter. Although some ADC architectures operate at fixed sampling rates, many practical digitization systems rely on tunable and phase-coherent clocks to support band selection, coherent undersampling, and multi-channel operation. A frequency-tunable, single-femtosecond microwave clock therefore offers a route to substantially reduce clock-induced sampling uncertainty while remaining compatible with flexible electronic ADC architectures~\cite{murmann2015}. Beyond ADC clocking, this capability opens new opportunities in electronic systems that require simultaneous spectral purity, frequency tunability, and absolute stability, including advanced communication, radar, spectroscopy, and time and frequency metrology.

More broadly, the availability of a frequency-tunable microwave source with single-femtosecond timing jitter represents a qualitative shift in what can be achieved in electronic systems. While ultralow jitter clocks have historically been accessible in the optical domain, their lack of flexibility has limited their use in practical electronic applications. The feed-forward eOFD approach provides a general pathway for dividing optical-frequency coherence into tunable electronic signals.

\section*{Methods}\label{sec:methods}
\subsection*{Feed-forward phase noise model}
To derive the phase noise model given by Eq. \ref{eq:pnmodel} we first consider the frequency domain output of the photodiode in Fig. \ref{fig:eOFDFF}\textbf{a},

\begin{equation}
    f_b = |f_0 - N f_m| + n_b,
\label{eq:fbeat}
\end{equation}

\noindent where $n_b$ is an additive noise term representing the finite signal-to-noise ratio of the photodiode signal. In practice, this term is dominated by the photodiode shot-noise, which is a function of the optical power. This term also includes any added noise from an amplifier (noise figure) that may follow the photodiode. The signal out of the frequency divider then divides Eq. \ref{eq:fbeat} by N, giving

\begin{equation}
    f_{div} = \frac{1}{N}|f_0 - f_m| + \frac{1}{N}n_b + n_{div},
\label{eq:fdiv}
\end{equation}

\noindent where $n_{div}$ is the residual noise of the frequency divider. Then, the output of the SSBM combines Eq. \ref{eq:fdiv} with another copy of $f_m$. If we assume no delay mismatch between inputs of the SSBM, the output is given by

\begin{align}
    f_{out} &= \frac{1}{N}|f_0 - f_m| + \frac{1}{N}n_b + n_{div} \pm f_m \nonumber\\
    f_{out} &= \frac{1}{N}f_0 + \frac{1}{N}n_b + n_{div},
\label{eq:fout}
\end{align}

\noindent where the relative phase (sign) between $f_b/N$ and $f_m$ can be chosen experimentally by the configuration of the SSBM (see next methods section). In practice, canceling the $f_m$ term leads to feed-forward compensation. We also treat the mixer as noiseless although in principle it could contribute to the thermal noise floor. In this study, the thermal noise was below the other additive sources of noise and did not need to be considered.

Now we can derive a phase noise PSD from Eq. \ref{eq:fout} by treating the frequency as fixed and assuming independent, time-varying phase fluctuations for each frequency component. This gives the output phase noise PSD

\begin{equation}
    S^{\mathrm{out}}_{\phi}(f) = \frac{1}{N^2}S^{0}_{\phi}(f) + \frac{1}{N^2}S^{b}_{\phi}(f)
    + S^{\mathrm{div}}_{\phi}(f).
\label{eq:pnidealmodel}
\end{equation}

\noindent Here, we can now show that if we kept the $2f_m$ term in eq. \ref{eq:fout}, we would have a corresponding phase noise term proportional to $S^{m}_{\phi}(f)$. Thus, choosing the correct sideband from the SSBM is a critical experimental detail.

To include the effects of the time delay, we consider the instantaneous phase of the microwave frequency signal when it meets at the SSBM

\begin{equation}
    \varphi_{\mathrm{residual}} = \varphi_m(t') - \varphi_m(t'').
\label{eq:residualphase}
\end{equation}

\noindent Here, $t'$ and $t''$ are the propagation time of the two copies of the microwave source signal. Taking the Fourier transform of \ref{eq:residualphase} gives 

\begin{equation}
    \Phi_{\mathrm{residual}}(f) = \left(1 - e^{-i 2\pi f(t'-t'')} \right) \Phi_m(f).
\label{eq:ft}
\end{equation}

\noindent where the capitals denote Fourier transformed functions. From this we can define a transfer function $|H_{\mathrm{ff}}(f)|^2$ such that 

\begin{align}
    |H_{\mathrm{ff}}(f)|^2 &= \left(1 - e^{-i 2\pi f(t'-t'')} \right)\left(1 - e^{+i 2\pi f(t'-t'')} \right), \nonumber \\
    &= 4\sin^2(\pi f\Delta\tau),
\end{align}

\noindent where $\Delta\tau = t' -t''$ is the delay mismatch between microwave paths. And the residual phase noise PSD from can then be written as

\begin{align}
    S^{m,\mathrm{res}}_{\phi}(f) &= |H_{\mathrm{ff}}(f)|^2 S^{m}_{\phi}(f)\nonumber\\
    &= 4\sin^2(\pi f\Delta\tau)~S^{m}_{\phi}(f).
\label{eq:psdresid}
\end{align}

\noindent Finally, adding Eq. \ref{eq:psdresid} to Eq. \ref{eq:pnidealmodel} results in the full phase noise model, Eq. \ref{eq:pnmodel}.

We note that Eq. \ref{eq:psdresid} shows an explicit dependence on the microwave source phase noise, $S^{m}_{\varphi}(f)$, which sets the limit to which the product $f\Delta\tau$ matters. The upshot is that an oscillator with large phase noise will require precise delay matching in order to maintain feed-forward cancellation at high Fourier frequency. We show this for three different oscillators at a fixed delay compensation of 50~ns in the Extended Data Fig. \ref{fig:sources}.

\subsection*{SSBM and output spectrum}
The single sideband mixer in the study was comprised of a 90-degree hybrid splitter/combiner and an IQ mixer. The IF port mentioned in the text comprised of two ports, I and Q, fed by the two outputs of the hybrid splitter. The sign of $f_0 - N f_m$ determines whether the upper or lower sideband contains the feed-forward phase information that will cancel upon combination. This was verified by swapping the I and Q inputs to the mixer. Selecting the ``wrong'' side-band doubles the noise of the original microwave source. Additional spectral components were suppressed by at least 20~dB by the SSBM. In practice, this was enough suppression for phase noise analysis. 

The Extended Data Fig. \ref{fig:spectrum} shows the typical output spectrum. The major contributions are the suppressed carrier and other sideband. Harmonics can also be observed. These spectral components are the dominant contributions to noise above 10~MHz Fourier frequency. Several methods can be used to further improve the spectrum. We observed good results from a narrow-band waveguide filter. The resulting spectrum is compared to the typical output and the resulting phase noise was measured. We also observed cross-collapse of the measured PSD near the filter band edges \cite{nelson2014}. While the waveguide filter provides excellent suppression of unwanted spectral elements, it is not tunable. Some tunable solutions include using a tunable YIG filter, or injection locking of a tunable oscillator. These were not immediately available for us to test and are left for future demonstrations.

\subsection*{Delay mismatch and compensation}\label{subsec:delay}

Varying lengths of coaxial cable were used to investigate the delay compensation required to mitigate phase noise contributions from the last term of Eq. \ref{eq:pnmodel}. Using the tunable microwave synthesizer (Keysight MXG) as a source, the phase noise PSD of the feed-forward output was measured for three different delays. The delay from each amount of cable was directly deduced from the light propagation speed in each cable allowing the delay mismatch to be determined by the phase noise model. The results of these tests are shown in Extended Data Fig. \ref{fig:timedelay}. A fixed delay length of $\approx10$~m was optimal for all experimental results in this paper.

When the feed-forward delay mismatch $\Delta\tau$ is compensated such that the residual microwave source contribution described by Eq.~\ref{eq:psdresid} is strongly suppressed, an additional excess phase-noise contribution is observed in the output PSD. This excess appears over the offset-frequency range from a few kHz to approximately 100~kHz and is not captured by the ideal feed-forward model (Eq. \ref{eq:pnmodel}).

We attribute this excess noise to amplitude-to-phase (AM-to-PM) conversion of the microwave signal at frequency $f_m$. In practice, AM-to-PM conversion can arise from distributed system effects, including, but not limited to, power-dependent phase response of the single-sideband mixer and changes in signal power associated with delay-compensation coaxial cabling and associated non-nonlinearities. As these contributions are difficult to isolate experimentally, we model AM-to-PM conversion phenomenologically.

The single-sideband fractional amplitude noise of the microwave source, $S_{\mathrm{AM}}(f)$, was measured independently and is shown in Extended Data Fig. \ref{fig:amtopm}. The spectral shape closely matches the observed excess phase noise with feed-forward cancellation. We model the resulting phase-noise contribution by assuming a linear conversion between fractional amplitude fluctuations and phase fluctuations, characterized by an effective AM-to-PM coefficient $k_{\mathrm{AM\to PM}}$ with units of radians per unit fractional amplitude. The corresponding phase-noise PSD is written as
\begin{equation}
S^{\mathrm{AM\to PM}}_{\phi}(f)
=
k_{\mathrm{AM\to PM}}^{2}~S_{\mathrm{AM}}(f).
\label{eq:ampm}
\end{equation}

Including this contribution, the total output phase-noise power spectral density becomes
\begin{align}
S^{\mathrm{out}}_{\phi}(f) &=
\frac{1}{N^2}S^{0}_{\phi}(f)
+
\frac{1}{N^2}S^{b}_{\phi}(f)
+
S^{\mathrm{div}}_{\phi}(f)
+
4\sin^2(\pi f \Delta\tau)~S^{m}_{\phi}(f)
\nonumber\\
&\quad+
S^{\mathrm{AM\to PM}}_{\phi}(f).
\label{eq:pnmodel_ampm}
\end{align}

This extended model quantitatively reproduces the measured output phase-noise spectra for both uncompensated and compensated delay conditions, as shown in Extended Data Fig. \ref{fig:amtopm}. In particular, when the feed-forward residual is minimized, the observed excess noise is well described by the AM-to-PM contribution using a single scaling parameter $k_{\mathrm{AM\to PM}}$. Over the measured offset-frequency range, the agreement between model and experiment indicates that treating $k_{\mathrm{AM\to PM}}$ as frequency independent is sufficient to capture the dominant AM-to-PM coupling in the present system. A detailed investigation of the origins of this conversion is beyond the scope of this work and will be addressed in future studies.

\backmatter
\newpage
\bmhead*{Acknowledgements}

The authors would like to thank John Dorighi and Keysight for use of the SSA-X phase noise analyzer used in this study. We also thank Hideyuki Ohtake and Yuki Ichikawa for their support.

\bmhead*{Competing interests}

The authors have no competing interests to declare.

\bmhead*{Data availability}
All of the underlying data in this study are available from the corresponding author upon reasonable request.

\bmhead*{Author contributions}
J.G., S.C.E., and B.M.H built the experiment. J.G. and S.C.E. measured and recorded the data. W.F.M, and A.R. helped with the data interpretation and analysis. All authors contributed to the manuscript. A.R. derived the quantitative model and supervised the study.

\bibliography{sn-bibliography}
\newpage

\renewcommand{\figurename}{Extended Data Fig.}
\renewcommand{\thefigure}{\arabic{figure}}
\setcounter{figure}{0}

\begin{figure}[htbp]
\centering
\includegraphics[width=0.95\textwidth]{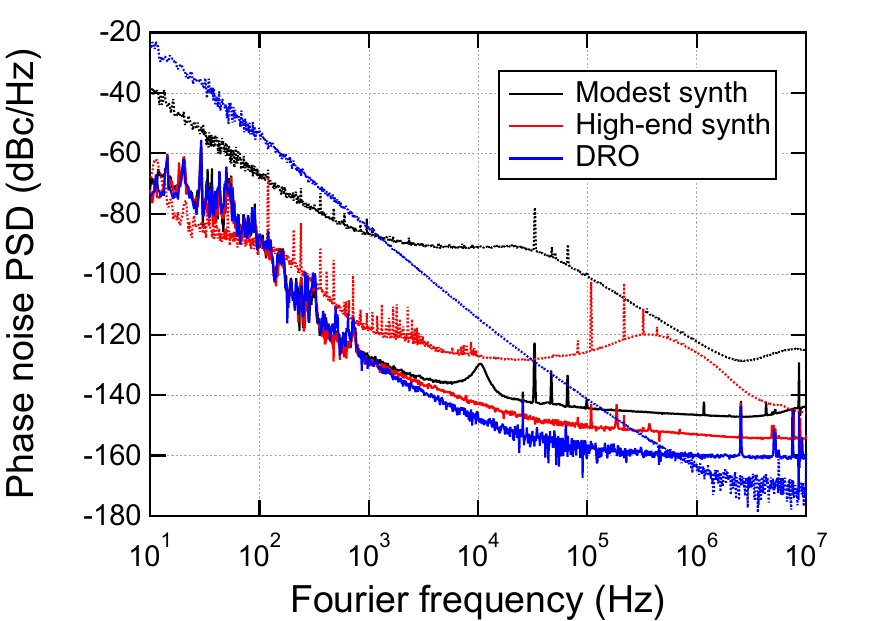}
\caption{Phase-noise power spectral density (PSD) of feed-forward eOFD outputs obtained using three different microwave sources under identical delay-compensation conditions. The dotted traces show the free-running phase noise of each microwave source, while the solid traces show the corresponding output phase noise after feed-forward eOFD. Results are shown for a modest-performance commercial synthesizer (black), a high-end commercial synthesizer (red), and a dielectric resonator oscillator (DRO, blue). Although the applied delay compensation is the same in all cases, the residual output phase noise at higher Fourier frequencies depends on the input microwave source phase noise $S^{m}_{\phi}(f)$, consistent with the delay-dependent term of the feed-forward phase-noise model. These data illustrate that, for a fixed delay mismatch, microwave sources with lower intrinsic phase noise enable deeper suppression and improved high-frequency performance. However, at lower Fourier frequencies, the optically divided reference performance is achieved independently of the microwave source phase noise.}\label{fig:sources}
\end{figure}

\begin{figure}[htp]
\centering
\includegraphics[width=0.8\textwidth]{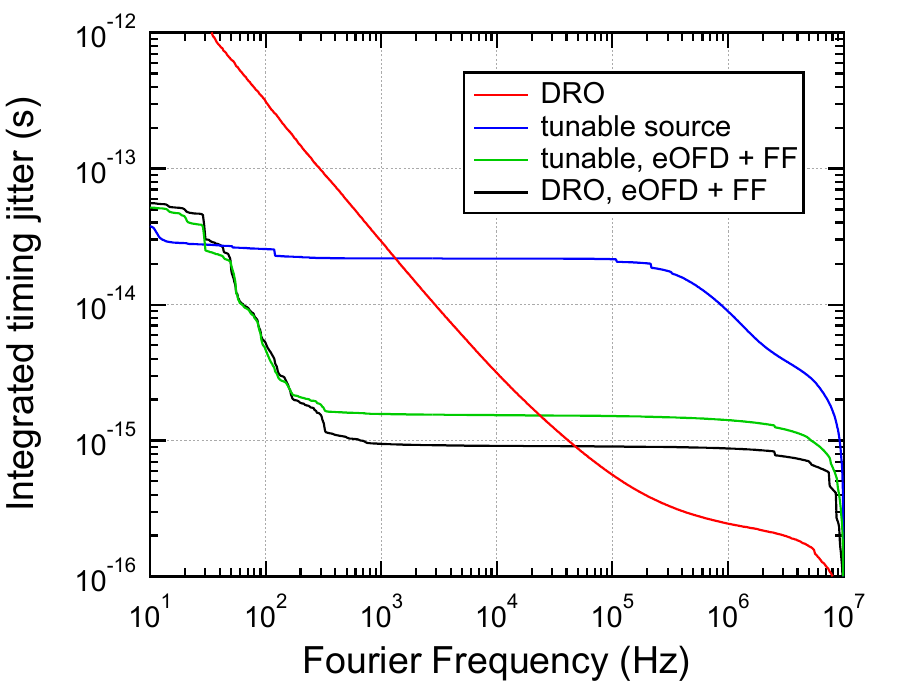}
\caption{Integrated timing jitter of the presented oscillators as a function of Fourier frequency. The integrated timing jitter is computed by integrating the measured phase-noise PSD from high to low Fourier frequency for 10~GHz outputs. Results are shown for the free-running DRO (red), tunable microwave source (blue), the synthesizer output after feed-forward eOFD (green), and the DRO-based eOFD output (black). At intermediate and high Fourier frequencies, feed-forward eOFD yields a substantial reduction in integrated timing jitter relative to the free-running synthesizer, reaching the single femtosecond level. Below approximately 300~Hz, drift of the dual-wavelength Brillouin laser (DWBL) optical reference contributes excess low-frequency noise, causing the integrated jitter to approach that of the commercial synthesizer. This low-frequency contribution is mitigated by external referencing of the optical splitting, as demonstrated in the main text.}\label{fig:inttj}
\end{figure}

\begin{figure}[htbp]
\centering
\includegraphics[width=0.75\textwidth]{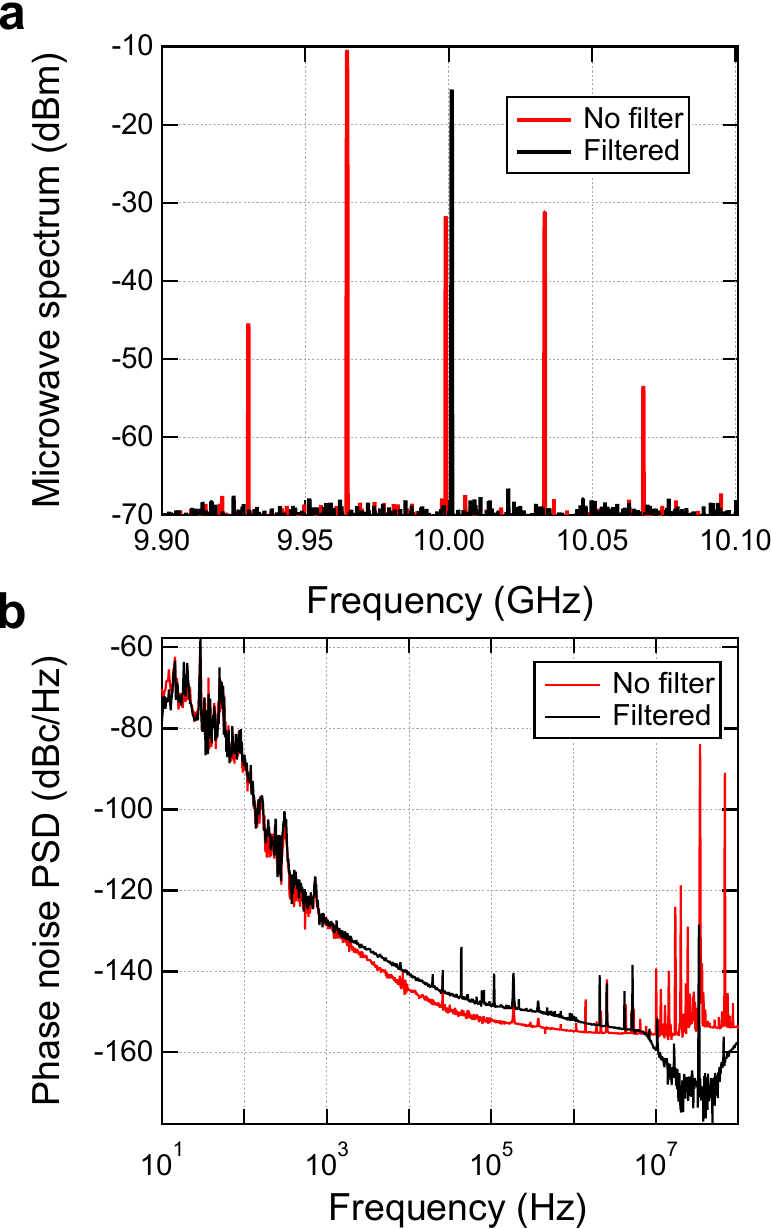}
\caption{\textbf{a} Microwave spectra of the output with and without filtering. The traces are offset in frequency by 0.1\% for clarity. Without a filter, the output is the lower sideband at $f_m - f_b/N = 9.97$~GHz. With the filter, the lower sideband output  was tuned exactly to $f_m - f_b/N = 10$~GHz to match the filter center frequency. \textbf{b} Phase noise PSD with and without the filter extending up to 100~MHz Fourier frequency. The marginal increase in phase noise for the filtered frequency output is caused by the frequency tuning mentioned above, without re-optimizing the SNR of $S^b_\varphi(f)$. In principle the phase noise should not degrade. Above 10~MHz Fourier frequency, the filter attenuates the output completely, leaving only thermal noise. This causes a cross-collapse in the phase noise spectrum, which is not a meaningful measurement of the true phase noise.}\label{fig:spectrum}
\end{figure}

\begin{figure}[htbp]
\centering
\includegraphics[width=0.95\textwidth]{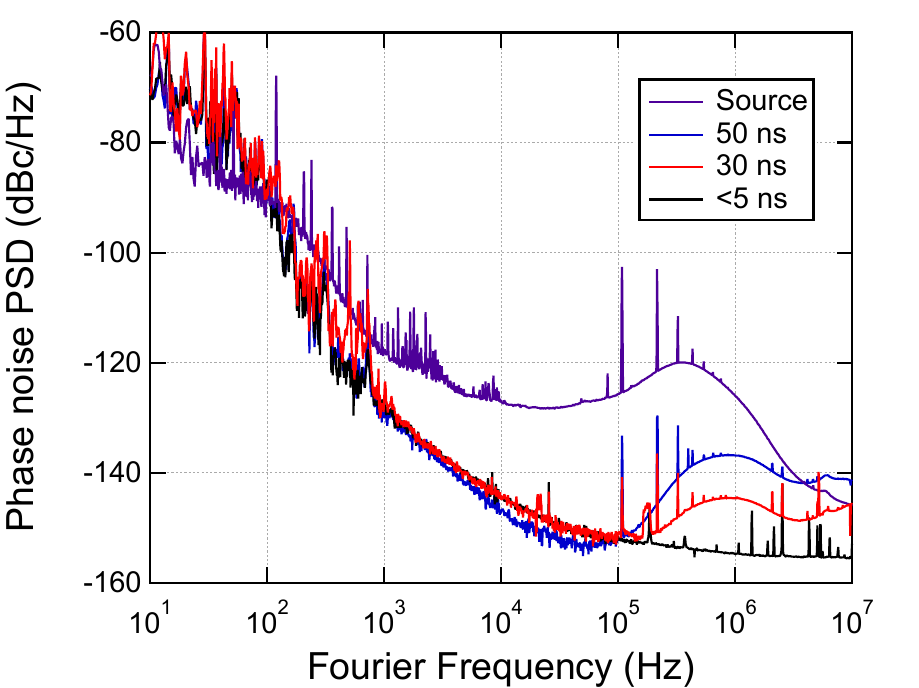}
\caption{Phase-noise power spectral density (PSD) of a 10~GHz microwave synthesizer after feed-forward eOFD using different amounts of delay compensation. The free-running synthesizer phase noise is shown for reference. The output phase noise is measured for three representative delay mismatches, $\Delta\tau \approx 50~\mathrm{ns}$ (blue), $\Delta\tau \approx 30~\mathrm{ns}$ (red), and $\Delta\tau < 5~\mathrm{ns}$ (black), estimated from the applied cable lengths and the phase-noise model. As the delay mismatch is reduced, the residual contribution from the microwave source phase noise is progressively suppressed, leading to improved phase-noise performance at higher Fourier frequencies. The data illustrate the sensitivity of feed-forward cancellation to path-length matching and validate the delay-dependent behavior predicted by the model.}\label{fig:timedelay}
\end{figure}

\begin{figure}[htbp]
\centering
\includegraphics[width=0.95\textwidth]{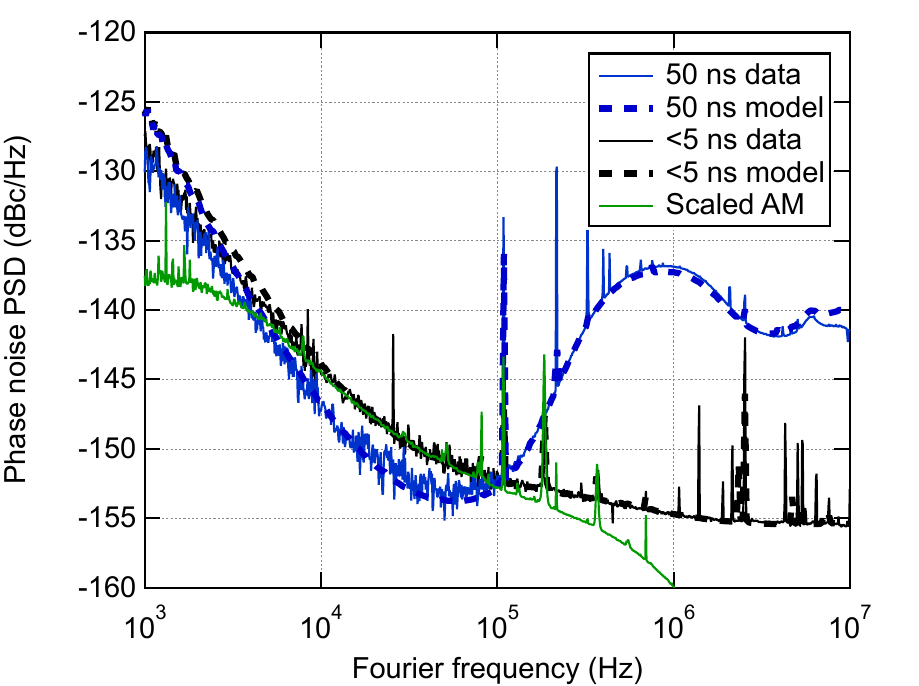}
\caption{Quantitative comparison between delay-compensated phase-noise measurements and the extended feed-forward phase-noise model including AM-to-PM conversion. Single-sideband phase-noise power spectral density (PSD) of the feed-forward eOFD output is shown for two representative delay-mismatch conditions, $\Delta\tau \approx 50~\mathrm{ns}$ (blue) and $\Delta\tau < 5~\mathrm{ns}$ (black), together with the corresponding model predictions (dashed lines). For the larger delay mismatch, the output phase noise exhibits the frequency-dependent degradation expected from incomplete feed-forward cancellation of the microwave source phase noise, in agreement with the $4\sin^2(\pi f \Delta\tau)S^{m}_{\phi}(f)$ term of the model. When the delay mismatch is reduced below $5~\mathrm{ns}$, the residual contribution from the microwave source is strongly suppressed, revealing an additional excess phase-noise contribution at intermediate Fourier frequencies. The green trace shows the independently measured single-sideband fractional amplitude noise of the microwave source, scaled by an effective AM-to-PM conversion coefficient and converted to an equivalent phase-noise PSD. The close agreement demonstrates that AM-to-PM conversion dominates the residual phase noise once delay-induced limitations are mitigated.}\label{fig:amtopm}
\end{figure}

\end{document}